\documentclass[
superscriptaddress,
 amsmath,amssymb,
 aps,
 pra,
 twocolumn,
 10pt
]{revtex4-2}

\usepackage[utf8]{inputenc}
\usepackage[T1]{fontenc}
\usepackage{graphicx}
\usepackage{array}
\usepackage{braket}
\usepackage[normalem]{ulem}
\usepackage{booktabs}
\usepackage{multirow}
\usepackage{color,soul}
\usepackage{mathtools}
\usepackage{hyperref}
\hypersetup{
	colorlinks,
	linkcolor={blue},
	citecolor={blue},
	urlcolor={blue}
}
\usepackage[capitalise]{cleveref}

\begin{document}

\title{A real-time, scalable, fast and highly resource efficient decoder for a quantum computer}
\author{Ben Barber}
\affiliation{Riverlane, Cambridge, UK.}
\author{Kenton M. Barnes}
\affiliation{Riverlane, Cambridge, UK.}
\author{Tomasz Bialas}
\affiliation{Riverlane, Cambridge, UK.}
\author{Okan Buğdaycı}
\affiliation{Riverlane, Cambridge, UK.}
\author{Earl T. Campbell}
\affiliation{Riverlane, Cambridge, UK.}
\affiliation{Department of Physics and Astronomy, University of Sheffield, UK}
\author{Neil I. Gillespie}
\affiliation{Riverlane, Cambridge, UK.}
\author{Kauser Johar}
\affiliation{Riverlane, Cambridge, UK.}
\author{Ram Rajan}
\affiliation{Riverlane, Cambridge, UK.}
\author{Adam W. Richardson}
\affiliation{Riverlane, Cambridge, UK.}
\author{Luka Skoric}
\affiliation{Riverlane, Cambridge, UK.}
\author{Canberk Topal}
\affiliation{Riverlane, Cambridge, UK.}
\author{Mark L. Turner}
\affiliation{Riverlane, Cambridge, UK.}
\author{Abbas B. Ziad}
\affiliation{Riverlane, Cambridge, UK.}

\date{September 2023}

\thispagestyle{plain}
\pagestyle{plain}
\begin{abstract}
    To unleash the potential of quantum computers, noise effects on qubits’ performance must be carefully managed. The decoders responsible for diagnosing noise-induced computational errors must use resources efficiently to enable scaling to large qubit counts and cryogenic operation. Additionally, they must operate at speed, to avoid an exponential slowdown in the logical clock rate of the quantum computer. To overcome such challenges, we introduce the Collision Clustering decoder and implement it on FPGA and ASIC hardware. We simulate logical memory experiments using the leading quantum error correction scheme, the surface code, and demonstrate MHz decoding speed – matching the requirements of fast-operating modalities such as superconducting qubits – up to an 881 and 1057 qubits surface code with the FPGA and ASIC, respectively. The ASIC design occupies 0.06mm$^2$ and consumes only 8mW of power. Our decoder is both highly performant and resource efficient, unlocking a viable path to practically realising fault-tolerant quantum computers.
\end{abstract}
\maketitle


\section{Introduction}\label{sec:intro}
Quantum computers could potentially solve computational problems that are out of reach of classical computers. However, to realise this potential all architectures need to deal with the fragility of their quantum bits (qubits) \cite{ftshor1996,ftAharonov1997,ftKitaev1997,ftKnill2005}. Qubits are highly likely to interact with the environment, leading to errors. Fortunately, Quantum Error Correction (QEC) protocols enable fault-tolerant computation in the presence of noise. These protocols are based on adding redundancy, encoding and protecting information into logical qubits by using a larger number of physical qubits. While errors can still corrupt the information, a signal is periodically generated from the logical data which characterises them. A decoder running on classical hardware processes this so-called syndrome, generating as an output the inferred error that has occurred, informing the corrective steps taken in subsequent operations. 

QEC must be performed continuously, creating a stream of syndrome data; as systems scale and logical error rates decrease, the amount of data that needs to be processed by a decoder increases significantly. Large computations will require real-time decoders that can process data at the rate it is received to avoid the creation of a backlog that grows exponentially with the depth of the computation~\cite{terhal2015Quantum,paralellwindow}, ultimately slowing it to a halt. Superconducting quantum devices, for example, generate a round of syndrome data in less than $1\mu$s (a rate of MHz), setting stringent requirements on decoder speed. Utility-scale quantum computers will require an optimised hardware decoder integrated in a tight loop at the heart of the control system.

Most experiments to date have used fast and accurate decoders implemented in software~\cite{higgott2023sparse,hig_bmbf2023,yue2023fusion} to decode offline~\cite{acharya2023SuppressingNature,delftqec2021,ETHQEC2022,InnsbQEC2022} rather than in real-time, the syndrome data being processed after the experiment has concluded. This type of decoding cannot support logic branching which is required to implement certain non-Clifford gates (the most essential gates to support quantum operations)\cite{Litinski2019gameofsurfacecodes}. Real-time decoding has been demonstrated in small scale experiments on ion-trap systems, using non-scalable lookup tables implemented in software that only require kHz speeds \cite{ryanA_qec21,ryanA_qec24}. However, in any scalable architecture, fast algorithmic decoders must be tightly integrated with the control system of the quantum computer to satisfy latency requirements. Decoders implemented on dedicated classical hardware, such as Field Programmable Gate Arrays (FPGAs) or Application Specific Integrated Circuits (ASICs), provide a viable path to such a solution.

To meet the challenge of developing real-time decoders, the community has begun to implement decoders on  FPGAs~\cite{lilliput, liyanage2023scalable, delftNN, Riste2020}, and provide models of implementations on ASICs ~\cite{delfosse_proc,delftNN}. FPGAs will be sufficient for the medium term. They provide the flexibility to adapt and change implementations of decoders, helping to identify the parameters needed to optimise the system performance. Until recently, only small instances of surface code decoders have been implemented on FPGAs~\cite{lilliput, Riste2020, delftNN}. Promising results have recently appeared on larger examples, where decoding an 881 qubit surface code memory simulation was demonstrated in under 1$\mu$s~\cite{liyanage2023scalable} per round. However, only a toy noise model was used and the design required significant FPGA resources.

FPGA systems have high per-unit cost and power consumption, hence they are not long-term solutions for scaling to millions of qubits. Cost-effective scaling of useful quantum computers will be achieved with ASICs, which guarantee improved performance and reduced power consumption at the cost of longer development times. Tight integration between decoders and control systems in a cryogenic environment will require ASICs~\cite{cryo-delft}.

In this work, we introduce the Collision Clustering (CC) decoder, designed to require few logical resources on an FPGA and low ASIC power and area occupation, while being performant enough to keep up with the syndrome generation time of the QPU. To demonstrate this, we implement CC on a Xilinx Ultrascale+ XCVU3P FPGA~\cite{xilinx_ultrascale}, and using industry leading EDA tools \cite{synopsys}, we also design an implementation on a 12nm FinFET ASIC process node, signed off to a standard that is ready to be taped out. Assuming a realistic circuit-level noise model, on the FPGA we decode an 881-qubit surface code in 810ns using only $4.5\%$ of the available computational elements (logic LUTs) and 10KB of memory. Moreover, we obtain a threshold of $0.78\%$. The ASIC decodes a 1057 qubit surface code in 240ns, using only 0.06mm$^2$ of area and 8mW of power.

The logical memory experiment simulated in this work preserves a state for a finite time period. To preserve a logical state indefinitely, the decoder must be able to handle data being streamed in. We save this investigation for future work.

\begin{figure}[ht]
    \centering
    \includegraphics[width=0.8\linewidth]{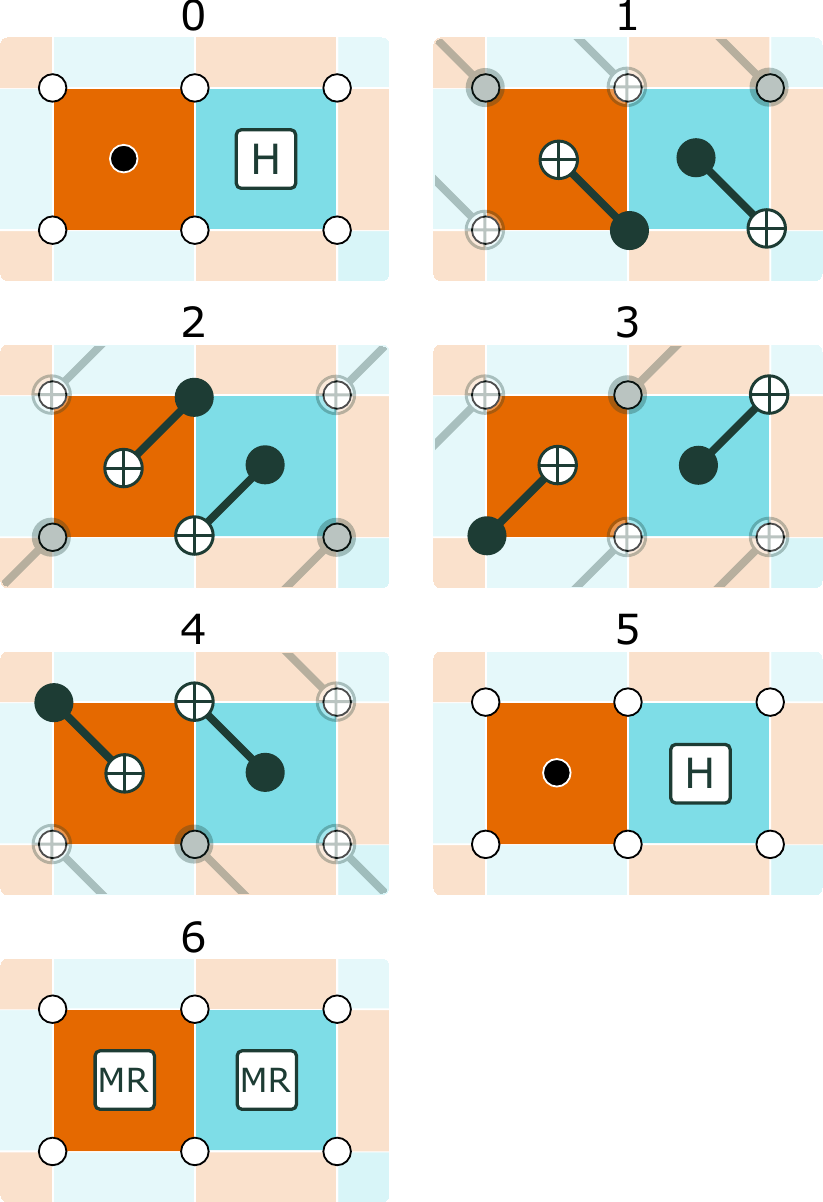}
    \caption{Syndrome extraction circuit for a section of rotated planar code. The circuit consists of Z (orange) and X (teal) stabilizer measurements that are preformed with 4 layers of CNOT gates and two layers of single-qubit (H) gates. Finally, the ancilla qubits are measured and reset (MR) for the next round. Errors can happen at any stage of the circuit, resulting in the circuit-level noise model described in detail in \cref{app:noise-model}}
    \label{fig:circuit-schedule}
\end{figure}

\section{The Collision Clustering decoder}
\label{sec:ccd-algo}

Collision Clustering (CC) is an implementation of Union-Find~\cite{Delfosse_2021,delfosse_proc}, a decoding algorithm with ``almost-linear'' asymptotic scaling. The input for CC is a decoding graph (\cref{app:decodesurface}). The vertices of this graph are all the possible defects that can occur when running the syndrome extraction circuit (\cref{fig:circuit-schedule}), where a \emph{defect} is a change in the measurement outcome of a syndrome qubit from one round of syndrome measurements to the next. The edges of the graph correspond to error mechanisms and are incident with the defects they cause. At a high level, CC decodes by first partitioning the set of defects of an input decoding graph into distinct subsets, known as clusters. Each cluster is then decoded separately using a simple procedure. 

Das et al.~described a detailed instruction set (micro-architecture) for implementing Union-Find on an FPGA or ASIC~\cite{delfosse_proc}. A key component of their micro-architecture is 
the spanning tree memory (STM), a data structure used in tracking the state of emerging clusters. By simulating their design we identified a bottleneck when reading from and writing into the STM, which significantly slowed down the execution of the algorithm. Therefore, we designed CC with a more memory efficient data structure to track the state of emerging clusters at the expense of asymptotic scaling. Using this architecture, on system sizes of practical interest, and larger than those modelled by Das et al.~, we achieved the necessary decoding speeds.

To decode each cluster, CC uses a reference logical operator. The size of a minimal logical operator is the \emph{distance} of a code, and is indicative of the number of errors a code can suppress. In the case of the distance $d$ surface code, defined on a $d\times d$ square lattice of data qubits and requiring $d^2-1$ syndrome qubits, we use the minimal length $d$ logical Pauli operator running along one of the boundaries of the lattice as the reference operator (\cref{app:surfacecodes}). To account for errors in the syndrome extraction circuit, certain operators are repeatedly measured giving the decoding graph a $3$D structure. In this setting, the minimal logical operator is identified with a $2$D boundary of the decoding graph, which we call the logical boundary. Each odd size cluster that touches this boundary flips the logical measurement, and so the correction bit returned by CC is the parity of the number of odd sized clusters that touch the logical boundary.

\subsection{Growth and Merge of Clusters}
In CC, each defect begins in its own cluster. The clusters then \emph{grow} in the decoding graph, and, if two clusters collide, that is overlap, they \emph{merge} to form one cluster. A cluster continues to grow so long as it has an odd number of defects within it, or until it touches one of the open boundaries of the decoding graph. Once all clusters have stopped growing, this growth-and-merge stage of the algorithm terminates.

We keep track of the growing clusters in the Cluster Growth Stack (CGS) data structure (\cref{fig:cc-decoder}a). Each entry of the CGS contains the following information:
\begin{enumerate}
    \item A \emph{vertex\_id} to represent which defect the entry is for. 
    \item A \emph{growth\_radius} to represent how far this defect has grown in the graph. 
    \item A \emph{valid\_bit}, set to 1 if this defect should be grown in the next round of growth, and set to 0 if the cluster containing it has stopped growing.
\end{enumerate}
In the CGS, clusters grow by updating the growth radius of the valid entries of the stack, requiring only a single read and write operation per defect. This is significantly more memory efficient than the STM data structure proposed by Das et al.~\cite{delfosse_proc}. For example, for a distance $d=15$ surface code on $449$ qubits, our decoder uses $80\%$ less memory despite being able to handle a noise model that requires more resources.

The Parent table (\cref{fig:cc-decoder}a) keeps track of which cluster a vertex belongs to. It has an address in memory for each defect, which holds the address of the corresponding parent defect. Upon initialisation, each defect is its own parent. Any entry with this property is called a \emph{root}. When two clusters merge, we set the root of one cluster equal to the root of the other cluster, the choice of which to update being arbitrary.  
The Merge unit (\cref{fig:cc-decoder}a) determines whether two clusters should merge. It takes each pair of defects (one defect from each cluster), calculates the distance between them in the decoding graph, and checks if the sum of their growths is greater than this distance. If it is, the two clusters merge. For a CGS with $s$ entries, this leads to $s^2/2$ \textit{collision} detection comparisons. 

Key to checking cluster mergers is efficient computation of the distance between two defects. We exploit the structure of the surface code to develop combinatorial functions that compute the distances without the need to traverse the graph. For a phenomenological noise model, this function consists of computing the Manhattan distance on a cubic lattice, with modifications to account for boundaries. Further modifications are required to account for extra space-time edges for the more realistic circuit-level noise model (\cref{app:coordinate}).


\section{Collision Clustering micro-architecture}
\begin{figure*}[ht]
    \centering
    \includegraphics[width=\linewidth]{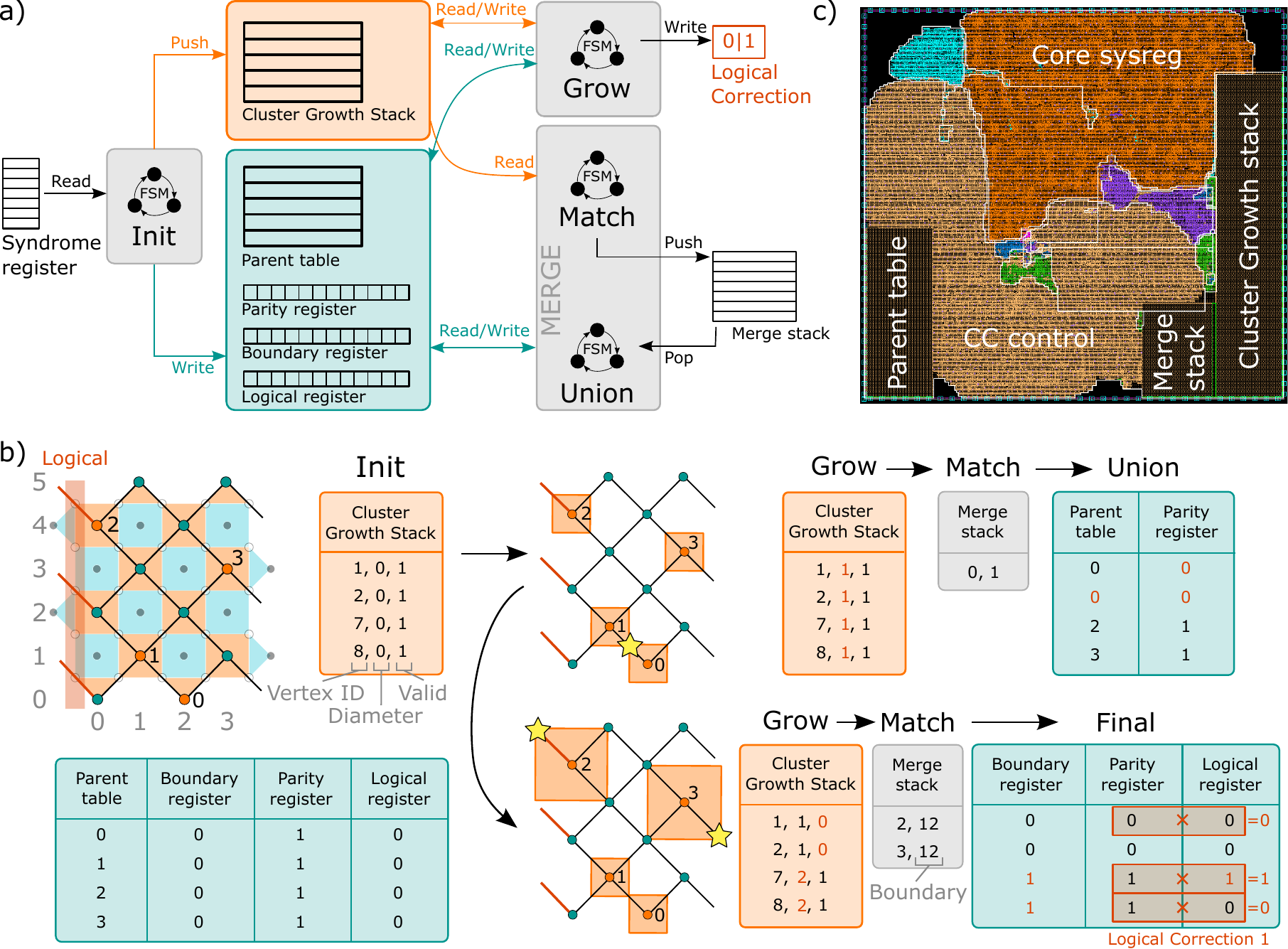}
    \caption{Collision clustering decoder.
    (a)~Micro-architecture diagram of the CC decoder computation engine, with annotated data flow. Each (sub-)unit uses a Finite State Machine (FSM) to control the computation. The input and output of the decoder is accessible through programmable registers. The logical correction bit is recalculated on the fly during Grow. The Merge processing element consists of Match and Union sub-units active in parallel.
    (b)~Schematic of the CC decoder on an example of a single round of a rotated planar surface code. In the Init image, the decoding graph with orange-highlighted defect vertices is laid on top of the surface code with orange and teal squares, which correspond to different parity check operators. First, in the Init step, the data structures are initialised with every defect being its own parent. In the main loop, the Grow step calculates the validity of the clusters and increases the diameter of all valid clusters. The clusters are checked for collisions in the Merge step and colliding pairs of clusters (highlighted by stars in the image) are pushed to the Merge stack. The Union step pops the pairs from the Merge stack, makes one cluster root the root of the other, and updates the parity, boundary and logical registers of the root of the cluster. Boundary and logical registers are only changed when merging with the boundary -- marked by auxiliary vertex 12 in the second growth step. By convention, the logical is defined to be the set of (orange) edges going to the boundary on the left side of the code and is changed to 1 when a cluster merges with the left boundary. When all clusters are invalidated, the final logical correction is calculated by summing logical registers of the roots of odd clusters.
    (c)~The floorplan of a distance $d=23$ (1057 qubits) ASIC implementing the CC decoder. Annotated: \emph{Parent Table}, \emph{Merge Stack} and \emph{Cluster Growth Stack} SRAM cells; \emph{CC control} logic formed of Init, Grow, and Merge processing units that are implemented as Finite State Machines (FSM). \emph{Core sysreg} logic contains input syndrome registers, control registers and the Metric Generation Unit. Other coloured regions contain clocking, IO and other miscellaneous logic.
    }
    \label{fig:cc-decoder}
\end{figure*}

Our micro-architecture of CC (\cref{fig:cc-decoder}a) is composed of shared memories and registers, and three processing units: Initialisation, Growth and Merge. The execution of the algorithm and the associated data structures are shown on a simple example in \cref{fig:cc-decoder}b.

A set of internal memories and registers keep the intermediary computational state. The growth of the clusters is tracked in the Cluster Growth Stack memory (\cref{fig:cc-decoder}a) as previously described. Recall also that the Parent Table (\cref{fig:cc-decoder}a) keeps track of the clusters; each defect is represented by an address in memory, and the data represents its corresponding parent defect. Three registers keep track of parameters for each growing cluster, the Boundary, Logical and Parity registers (\cref{fig:cc-decoder}a). In each, a defect is represented by an address. For the Boundary and Logical registers, the bit is set to $1$ if the corresponding cluster touches any boundary, respectively the logical boundary. In the Parity register, the bit is set to $1$ if there are an odd number of defects in the cluster.

Upon initialisation of the decoder, the \emph{Init} unit (\cref{fig:cc-decoder}a) processes the decoder configuration, loads the input syndrome data and appropriate data into the storage elements. 

The \emph{Grow} unit (\cref{fig:cc-decoder}a) is responsible for growing clusters. It updates the Cluster Growth Stack cluster entries at every iteration, finding the root of a cluster, then writing the radius and validity status. While processing a cluster, the Grow unit also checks for collisions with either boundary, writing the colliding vertex and the boundary or logical vertex to the Merge stack. The Grow unit also simultaneously computes the logical correction bit resulting from the growth stage. The correction is discarded and recomputed on the next growth cycle if growth is required, or is kept and used as the output correction.

The \emph{Merge} unit consists of \emph{Match} and \emph{Union} sub-units (\cref{fig:cc-decoder}a), operating in parallel. The Match sub-unit performs collision detection comparisons from CGS data. It then writes colliding defect pairs onto the Merge stack. The Union sub-unit reads collided vertex entries from the Merge stack, then searches the Parent Table for the two roots. It then updates the Parent Table, Logical, Boundary and Parity registers with the results of the clusters' union.

\section{Physical implementation on FPGA}

\begin{figure*}[ht]
    \centering
    \includegraphics[width=\linewidth]{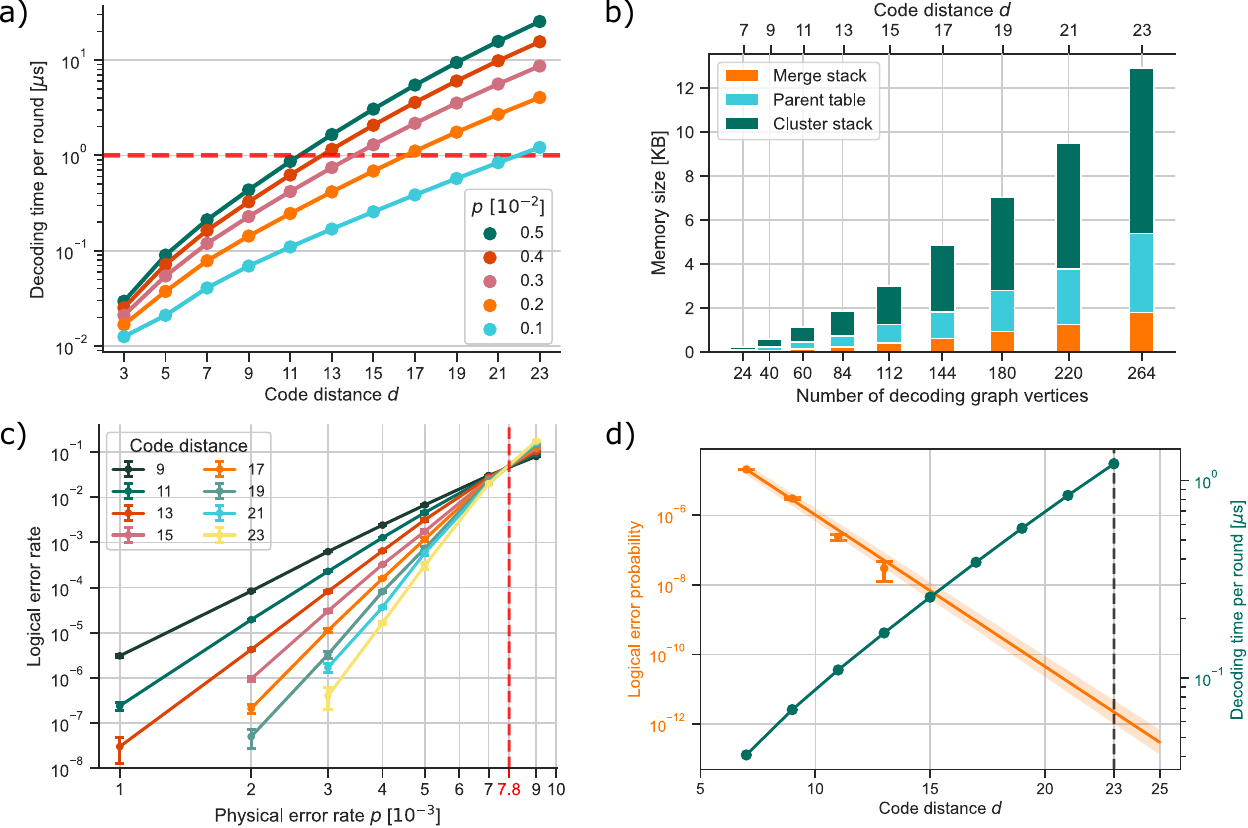}
    \caption{Performance of our FPGA (Xilinx Ultrascale+ XCVU3P~\cite{xilinx_ultrascale}) implementation of the CC decoder on the rotated planar surface code.
    (a)~Decoding time per syndrome measurement round as a function of code distance for different noise rates. Even at a high noise just below threshold ($p=0.5\%$) expected in near-term hardware, our decoder can decode large ($d=11$) codes at sub-1$\mu$s per round rate. 
    (b)~FPGA memory size usage of different data structures for varying code distances using $p=0.1\%$. Even at distance $d=23$ the decoder uses below 13KB of memory, allowing for implementation on affordable FPGA hardware.
    (c)~Logical error probability for varying physical error rate $p$ and a range of code distances demonstrating threshold at around $p=7.8\times 10^{-3}$. Points are generated using a maximum of $10^8$ samples and the points with no more than 1 error are removed.
    (d)~Decoding time per round at $p=0.1\%$ together with a projection of logical error probability to large distance regime. The decoder is just short of 1$\mu$s per round at distance $d=23$ on the affordable Xilinx Ultrascale+ XCVU3P hardware while expected logical error probability is approaching $10^{-12}$. Each accuracy point was obtained using $10^8$ samples. Dashed vertical line is a guide to the eye. Orange line is an exponential decay fit to the data and the orange shadowed region is the standard deviation to the projection. All error bars are the standard error of the mean.
    }
    \label{fig:fpga-performance}
\end{figure*}

In Table~\ref{tab:FPGA-results-rplanar} we present the performance of our FPGA implementation of CC applied to the rotated planar surface code across a range of distances from $3$ ($17$ qubits) to $23$ ($1057$ qubits). We use a circuit level noise model (\cref{app:noise-model}), specifically a depolarization channel after each $2$-qubit gate with probability $p$, and a depolarization channel after each $1$-qubit gate, measurement and reset, each with probability $p/10$. We also flip the outcome of each measurement with probability $p$. For the data in Table~\ref{tab:FPGA-results-rplanar}, we set $p=0.1\%$. We also plot in \cref{fig:fpga-performance}a the decoding time per round for varying distances and values of $p$. We see that the FPGA decoder can decode below the $1\mu$s threshold up to distance $21$. We targeted a maximum clock frequency (FMax) of 400 MHz, and due to the low resource utilisation, FMax is not significantly impacted by increasing the distance of the code.

The precise resource requirements on the FPGA, including the percentage of resource utilised, are also given in Table~\ref{tab:FPGA-results-rplanar}. The only resources required to implement CC are trivial logic gates along with storage elements. Notably, no Digital Signal Processing (DSP) elements are needed, whose use would increase the area of the implementation. 

One of the main advantages of CC is its efficient use of storage resources. \cref{fig:fpga-performance}b shows, for each code distance, the storage required for the main data structures. The micro-architecture has other storage elements e.g.~Boundary, Logical and Parity registers. These are typically implemented as flip-flops and the overall results are given in Table~\ref{tab:FPGA-results-rplanar}. For a distance $23$ implementation, the main storage requirement is around 13KB which is a third of a level-1 data cache size in a typical application CPU. Similar to these caches on CPUs, the CC memories can be accessed at very high frequencies in a single clock cycle, enabling performant data processing.

Some of the other known FPGA based hardware decoders~\cite{lilliput, liyanage2023scalable} require substantial resources at larger code distances. Their data structures are also sized based on a phenomenological noise model at $p=0.1\%$, so they have significantly under-counted resources compared to the requirements for a circuit-level noise model of the same magnitude, a more realistic noise model (\cref{app:noise-model}).

\begin{table*}[ht]
\centering
\begin{tabular}{@{}cccrrrrr@{}}
\toprule
\multicolumn{1}{c}{\multirow{3}{*}{\textbf{\shortstack{\\ Code\\ Distance}}}} &
  \multicolumn{2}{c}{\textbf{FPGA Performance}} &
  \multicolumn{5}{c}{\textbf{FPGA Utilisation}} \\ \cmidrule{2-8} 
\multicolumn{1}{c}{} &
  \multicolumn{1}{c}{\textbf{\shortstack{Fmax\\ \lbrack MHz\rbrack}}} &
  \multicolumn{1}{c}{\textbf{\shortstack{Exec-time\\ \lbrack µs\rbrack}}} &
  \multicolumn{1}{c}{\textbf{Logic LUTs}} &
  \multicolumn{1}{c}{\textbf{LUTRAMs}} &
  \multicolumn{1}{c}{\textbf{FlipFlops}} &
  \multicolumn{1}{c}{\textbf{RAMB36}} &
  \multicolumn{1}{c}{\textbf{RAMB18}} \\  \cmidrule{1-8}
3  & 449 & 0.07 & 2491 (0.63\%)  & 12 (0.01\%)  & 1572 (0.20\%)  & 0 (0.00\%) & 0 (0.00\%) \\
5  & 445 & 0.06 & 2709 (0.69\%)  & 27 (0.01\%)  & 1670 (0.21\%)  & 0 (0.00\%) & 0 (0.00\%) \\
7  & 406 & 0.06 & 3092 (0.78\%)  & 49 (0.02\%)  & 1984 (0.25\%)  & 0 (0.00\%) & 0 (0.00\%) \\
9  & 408 & 0.07 & 3800 (0.96\%)  & 100 (0.05\%) & 2483 (0.32\%)  & 0 (0.00\%) & 0 (0.00\%) \\
11 & 412 & 0.11 & 4600 (1.17\%)  & 68 (0.03\%)  & 3591 (0.40\%)  & 0 (0.00\%) & 1 (0.07\%) \\
13 & 411 & 0.16 & 5914 (1.50\%)  & 105 (0.05\%) & 4136 (0.52\%)  & 0 (0.00\%) & 1 (0.07\%) \\
15 & 403 & 0.25 & 7793 (1.96\%)  & 60 (0.03\%)  & 5432 (0.69\%)  & 1 (0.14\%) & 1 (0.07\%) \\
17 & 408 & 0.37 & 10446 (2.66\%) & 90 (0.05\%)  & 7184 (0.91\%)  & 1 (0.14\%) & 1 (0.07\%) \\
19 & 402 & 0.55 & 13331 (3.38\%) & 0 (0.00\%)   & 9277 (1.18\%)  & 2 (0.28\%) & 2 (0.14\%) \\
21 & 405 & 0.81 & 17237 (4.37\%) & 0 (0.00\%)   & 11957 (1.52\%) & 2 (0.28\%) & 2 (0.14\%) \\
23 & 401 & 1.18 & 21693 (5.50\%) & 0 (0.00\%)   & 15126 (1.92\%) & 5 (0.69\%) & 1 (0.07\%) \\ 
\hline
\end{tabular}%
\caption{FPGA results for decoding the rotated planar surface code, assuming $p=0.1\%$. 
Code distance: the size of the error correcting code. 
Fmax: the maximum achieved clock frequency on the targeted FPGA. 
Exec-time: the execution time averaged over 100,000 shots, normalised by dividing by the $d$ rounds of syndrome generation. 
Logic LUTs: the number of FPGA lookup tables used for logic primitives. 
LUTRAMs, FF, RAMB36 and RAMB18: different types of storage elements. The mapping of CC data structures to a storage element depend on the size of the structure. The numbers in round bracket are the percentage of the corresponding type of resource used on the FPGA.}
\label{tab:FPGA-results-rplanar}
\end{table*}

As well as being performant and resource efficient, our implementation needs to be accurate to effectively suppress physical errors. The clusters generated by CC are the same as those generated by Union-Find, resulting in the same accuracy. This intuitively holds since any cluster in Union-Find is the union of balls of different radii centered on the defects, and is confirmed empirically. We demonstrate the accuracy of CC by calculating a threshold plot \cite{fowler_threshold_2009}, given in \cref{fig:fpga-performance}c. Our implementation has a threshold of $0.78\%$, which means that for values of $p$ lower than this threshold errors are suppressed exponentially by increasing the code distance. To further confirm the accuracy of CC, in \cref{fig:fpga-performance}d we estimate the distances required to obtain very small logical error rates using CC. We first calculate the logical error rate for code distances up to $d=13$ using $10^8$ shots per data point. The resulting small error bars enable us to accurately project out to the small logical error rate regime. We see that using CC, we can obtain a logical error rate approaching $10^{-12}$ with only a distance $23$ surface code.

\section{Physical Implementation on ASIC}
Our ASIC implementation design is ready to be taped out, having been signed-off using industry-leading Electronic Design Automation (EDA) tooling \cite{synopsys}, top-tier foundry silicon-proven multi-Vt libraries, SRAM IP and spice models. This ensures high quality results which include all the fabrication process variations of device models and parasitic effects from the power network, clock-tree synthesis, place and route stages.

In Table~\ref{tab:ASIC-results} we present two physical implementations of CC on a 12nm FinFET process node: decoding a distance $7$ and a distance $23$ surface code. They respectively take $10$ns and $240$ns to decode per round of syndrome measurements, using only $2.75$mW and $7.85$mW of power. 

The control systems used today for quantum computers cannot be scaled to control the large numbers of qubits needed to run QEC schemes that obtain low logical error rates. Cryogenic CMOS based control systems~\cite{cryo-ibm, cryo-google, Pauka2021} could represent a solution, in which case only tight integration of the decoder will lead to optimal performance. Current cryogenic systems however have a strict power budget, in the order of 1W at the 4K temperature range~\cite{krinner2019engineeringa}. We envisage a maximum power budget of tens of mW for a decoder, with qubit control and readout remaining the primary consumption sources. In addition to our ASIC implementations satisfying these power budget constraints, in the near term our distance $7$ instance of CC will be valuable in testing error correction experiments using cryogenic CMOS based control systems.

\begin{table*}[hbt!]
\centering
\begin{tabular}{@{}ccccccc@{}}
\toprule
\multicolumn{1}{c}{\multirow{3}{*}{\textbf{\shortstack{\\ Code\\ Distance}}}}
 & \multicolumn{2}{c}{\textbf{ASIC Performance}}                         
 & \multicolumn{2}{c}{\textbf{ASIC Area}} 
 & \multicolumn{2}{c}{\textbf{ASIC Power}} 
 \\ \cmidrule(lr){2-7}
 & \multicolumn{1}{c}{\textbf{\shortstack{Fmax\\ \lbrack MHz\rbrack}}}
 & \multicolumn{1}{c}{\textbf{\shortstack{Exec-time\\ \lbrack µs\rbrack}}}
 & \multicolumn{1}{c}{\textbf{\shortstack{Die-size\\ \lbrack mm\textsuperscript{2}\rbrack}}}
 & \multicolumn{1}{c}{\textbf{FlipFlops}}
 & \multicolumn{1}{c}{\textbf{\shortstack{Dynamic\\ \lbrack mW\rbrack}}}
 & \multicolumn{1}{c}{\textbf{\shortstack{Leakage\\ \lbrack mW\rbrack}}}

 \\
  \cmidrule{1-7}

7 & 2000 & 0.01 & 0.009 & 3957 & 2.73 & 0.02 \\
23 & 2000 & 0.24 & 0.064 & 15840 & 7.72 & 0.13\\
\cmidrule{1-7}
\end{tabular}
\caption{\label{tab:ASIC-results}CC ASIC results for decoding the rotated planar surface code using a 12nm FinFET process assuming p=0.1\%. We use industry-leading EDA tools to determine the frequency of the implementation. To calculate the execution time, we use this frequency along with the cycle counts of the FPGA implementations.}
\end{table*}

A floorplan defines the approximate locations, sizes and shapes of various logical blocks of the design. It helps determine how signals will interact between different blocks, enabling performance, power and area optimization. The floorplan for a distance $23$ implementation is shown in \cref{fig:cc-decoder}c. 


\section{Discussion}

\begin{figure}
    \centering
    \includegraphics[width=\linewidth]{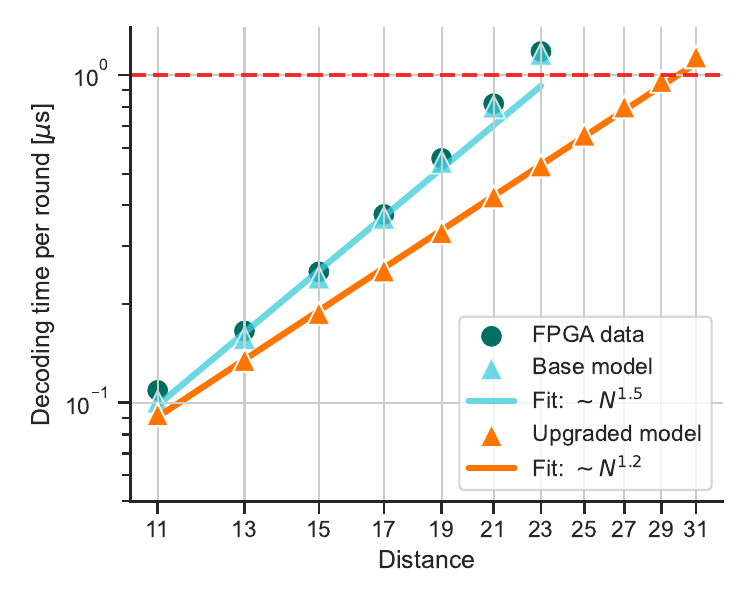}
    \caption{Modelling improvements to the CC algorithm for the next generation FPGA decoder with $p=0.1\%$. The results of our in-house modelling tool (base model, teal) are compared with the FPGA-acquired data (green) from \cref{fig:fpga-performance}d showing a high degree of correlation. The fit to the data is showing a $\sim N^{1.5}$ scaling as the asymptotic regime has not been reached. The modelling of the improvements to the algorithm (upgraded model, orange) demonstrates a $\sim N^{1.2}$ scaling for the whole range of distances, allowing us to stay below 1$\mu$s/round threshold up to $d=29$. We assume 400 MHz FPGA for the modelled data.}
    \label{fig:hwi-modelling}
\end{figure}

To the best of our knowledge, there have been four demonstrations of decoders implemented on dedicated classical hardware~\cite{lilliput,liyanage2023scalable,delftNN,Riste2020}. Lookup table decoders are implemented on FPGAs in~\cite{lilliput} and~\cite{Riste2020}. In both cases, the error correction scheme demonstrated is relatively simple; the distance $3$ repetition code~\cite{Riste2020}, and the distance $5$ surface code~\cite{lilliput}. The exponentially scaling memory requirements make lookup table decoders impractical for surface code distances above $5$. Contrary to this, we have demonstrated that the CC decoder can easily scale to handle surface code distances of practical interest.

In~\cite{delftNN}, a neural network surface code decoder is implemented on an FPGA, up to only distance 5. Measurement errors are not considered, limiting its effectiveness and understanding of how the decoder will perform with experimental qubits, counter to the more realistic noise model we use. Additionally, the corresponding estimated performance, power and area of the ASIC synthesis all degrade significantly when increasing the distance from $3$ to $5$, implying that the design will not effectively scale, again in contrast to the efficient scaling of CC. 

The most significant prior implementation of a surface code decoder on classical hardware is found in~\cite{liyanage2023scalable}. A highly distributed implementation of Union-Find, called Helios, is implemented on an FPGA, assuming a phenomenological noise model. Each vertex of the decoding graph is assigned a processing unit, and communication across the design is limited e.g.~typically only nearest neighbour processing units can communicate with each other. In this setting, a round of syndrome measurements for a distance $21$ surface code is decoded in $11.5$ns on an FPGA.  

The fundamental difference between Helios and CC is the amount of parallelisation that the Helios architecture can take advantage of. In Helios, all clusters are worked on simultaneously, whereas in CC, there is a limit to the amount of parallelisation that can be leveraged for performance. A consequence of this is that Helios can achieve faster decoding speeds. However, Helios' distributed network requires data transfer between neighbouring processing elements, which will always require resources. CC uses a global data structure e.g.~the distance function, and optimisations of this global data structure can minimize resources. This difference is manifest when comparing the two implementations. Helios requires a large numbers of lookup tables (LUTs) and registers (900k LUTs and 240k registers). Registers are necessary in a distributed implementation so that data can be accessed at the same time in a small number of cycles. However, a register is an order of magnitude bigger in size and power consumption compared to a bit in memory. Hence, the numbers of LUTs and registers required by Helios will lead to a large area and power consumption, increasing the cost of any chip developed and preventing the integration with a cryogenic control system. Our implementation of CC is very resource efficient (Table \ref{tab:FPGA-results-rplanar}) while at the same time satisfies the performance requirements needed for superconducting qubits, and so is amenable to operating in cryogenic environments (Table \ref{tab:ASIC-results}).

Although we have already demonstrated CC decoding at speed with low FPGA utilization for large distance codes, we are developing further improvements to enhance its performance in future generations. Currently, the Match stage performs all-to-all cluster collision checks at every growth step, a bottleneck at large distances. We can remedy this by first taking advantage of the syndrome ordering in time, reducing the number of comparisons. Secondly, as the clusters get invalidated in the later iterations of the Grow-Merge loop, most comparisons are between invalid clusters. We can avoid this by keeping track of the valid clusters and ensuring that the comparisons between invalid clusters are removed.

To quantitatively assess the impact of such improvements, we modelled the enhanced CC algorithm using a hardware-indicative Python library which predicts the number of FPGA cycles and memory footprint. This model has been successfully validated using experimental data based on the current CC algorithm implementation (\cref{fig:hwi-modelling}), giving us confidence in our projections. The enhanced CC algorithm is expected to improve the scaling to $\sim N^{1.2}$; as a result, we will be able to decode a distance 29 surface code in under 1$\mu$s with only a modest sized FPGA. This is a step technological improvement with respect to \cite{liyanage2023scalable}, where the possibility to decode a distance 29 surface code in under 1$\mu$s was suggested (although not modelled), yet it would have required one of the largest commercially available FPGAs. 

\section{Conclusion}
In this work we introduced the Collision Clustering (CC) decoding algorithm and described a micro-architecture for its implementation. Fault-tolerant quantum computing requires a decoder to process error syndromes at speed in order to prevent a decoding backlog that exponentially slows down the logical clock rate. Moreover, any scalable quantum computer requires a decoder to be resource efficient, which will also enable tight integration with control systems in a cryogenic environment. To meet these requirements, we designed CC to be memory and power efficient. While CC has non-linear asymptotic scaling, this is remedied with parallelisation and pipelining for the relevant code distances, demonstrating that CC is a scalable, fast and highly resource-efficient decoder.

To verify this, we implemented CC on both an FPGA and ASIC. We decoded a logical memory experiment using large distance surface code examples in under 1$\mu$s per syndrome measurement round assuming a circuit-level noise model. On a modest sized FPGA, a distance 21 surface code took $810$ns to decode per round, utilising only $4.5\%$ of the available resources, and on a $12$nm FinFET process node, a distance $23$ surface code took $210$ns to decode per round using only $0.06$mm$^2$ area and $8$mW power.

To preserve a logical state indefinitely, \emph{sliding window decoding} \cite{dennisetal08} can be used. While continuous rounds of syndrome measurements are being generated, the decoder processes only a contiguous set, or \emph{window}, of these rounds. Utilising the whole window, the decoder commits to a correction for the longer lived defects in the window, storing it in software. The window then slides up to include more recent rounds of measurements, the process repeats, and the correction is updated. Certain boundary effects make this process more complex than the one simulated in this work. Therefore, developing a fast and efficient sliding window implementation of CC will be an important next step in the advancement of decoders for fault-tolerant computation.

\section*{Data Availability}

The stim~\cite{gidney2021Stim} circuits used to generate the samples and raw data from all the plots in this study are available on Zenodo with the DOI identifier \href{https://doi.org/10.5281/zenodo.11621878}{10.5281/zenodo.11621877}.

\section{Acknowledgements}
We thank Steve Brierley and Jake Taylor for encouraging this research and related discussions. We also thank Maria Maragkou and Luigi Martiradonna for feedback on the manuscript.


\appendix

\begin{figure*}[ht]
    \centering
    \includegraphics[width=0.9\linewidth]{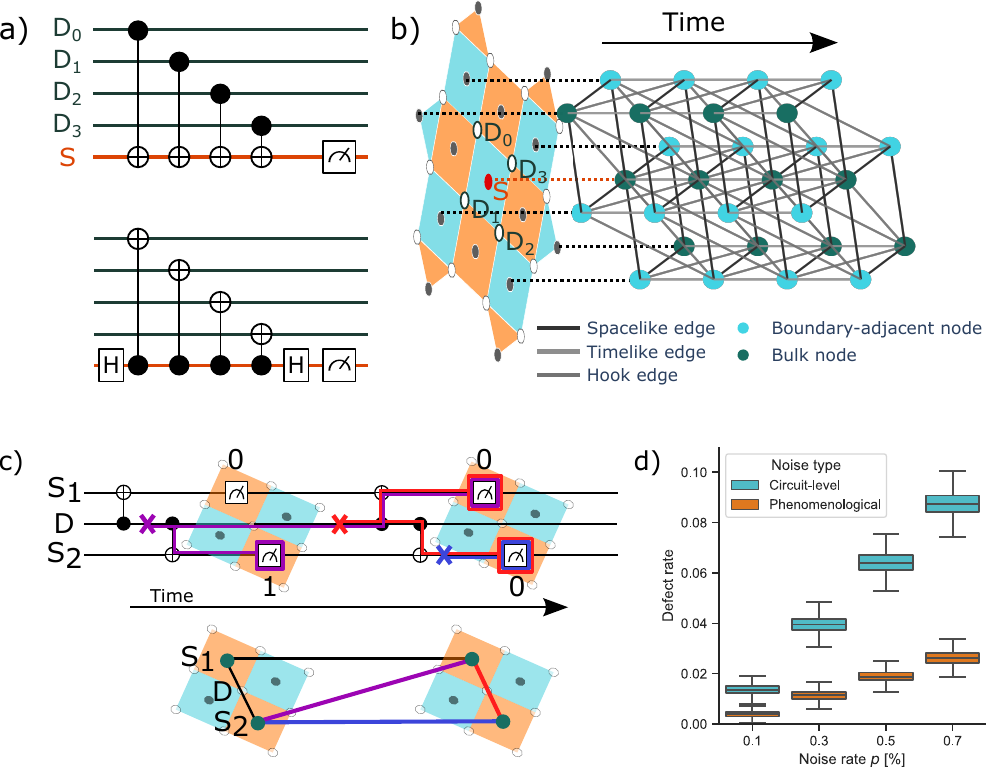}
    \caption{
    Quantum Error Correction using a rotated planar surface code.
    (a)~Quantum circuits to measure the Z (top) and X (bottom) parity checks. The circuit is continuously repeated until the computation ends and all qubits are measured out.
    (b)~The logical information is encoded by combining multiple physical qubits laid out on a surface (orange and teal grid). The physical qubits can be divided into data qubits ($D_i$, empty circles) encoding the state, and syndrome qubits ($S$, full circles). Repeated measurements of syndrome qubits provides information about errors. Possible error mechanisms can be represented by a graph in which nodes represent differences between syndrome measurements in consecutive rounds (potential defects) and edges represent error mechanisms that create the corresponding pair of defects (\cref{app:decodesurface} for more details). 
    Error mechanisms that trigger only one defect (teal nodes) are in addition connected to a fictitious boundary node (not drawn).
    Z and X parity check patches (teal and orange squares) protect against X and Z errors respectively, and we decompose any $Y$ error into an $X$ error and $Z$ error. The resulting disjoint decoding graphs are used to decode $X$ and $Z$ errors -- only the Z check graph is shown here.
    (c)~Two rounds of syndrome measurements displaying only a part of the circuit that involves a single data and two syndrome qubits (top). The possible error mechanisms are depicted with different colours: data (red), measurement (blue), and hook (purple) errors and result in corresponding edges in the decoding graph (bottom).
    (d)~Comparison of defect rate for phenomenological and circuit-level noise models with the same noise rate on distance 23 rotated planar surface code. While the decoding of circuit-level noise is made more complicated by the presence of hook edges, it also results in approximately 3.5 times more defects due to more possible error locations.
    }
    \label{fig:qec-cycle}
\end{figure*}

\section{Surface codes}
\label{app:surfacecodes}

Surface codes are a family of codes that have been studied extensively since their discovery over twenty years ago~\cite{kitaev2003fault}. They can be implemented on 2D architectures with fixed nearest neighbour interactions, a topology often used in QPUs based on superconducting qubits, for example. Moreover, extensive theoretical work has been developed to execute fault-tolerant computations based on these codes ~\cite{ChamberlandlatticePRX,fowlerlattice2018,Horsman2012,Litinski2019gameofsurfacecodes}. 
Surface codes also achieve the highest thresholds – corresponding to the minimum qubit physical error rates required to start correcting errors effectively – among currently available error correction schemes~\cite{fowler_threshold_2009}, and they have been implemented in several near-term error correction experiments~\cite{acharya2023SuppressingNature,delftqec2021,ETHQEC2022}. Combined, these observations make surface codes likely candidates for the error correction schemes used in the first fault-tolerant devices. In this work, we use the rotated surface code (\cref{fig:qec-cycle})~\cite{rotatedsurfacecode}.

The surface code is defined on a $d\times d$ square lattice (where $d$ is the number of data qubits in each dimension of the lattice) by operators that check the parity of sets of qubits in either the Pauli $Z$ or $X$ basis (\cref{fig:qec-cycle}). These operators are measured repeatedly to generate the syndrome and project the qubits into a logical computation space. We call a single round of measuring all the parity check operators a round of syndrome measurements, and the measurement data generated a round of syndrome data. For the rotated planar code, a round of syndrome measurements requires $d^2-1$ syndrome qubits, giving a total of $2d^2-1$ qubits. The logical qubit is defined by logical Pauli operators forming a path between opposite boundaries. The \emph{distance} of the surface code is the minimum number of Pauli operators in such a logical operator, which is just the side length $d$ of the lattice.

The quality of the QPU determines the initial error rate of the physical qubits. The distance of the surface code is a measure of the capability of this code to suppress the physical error rate down to a target logical error rate – the larger the distance, the lower the logical error rate. Quantum algorithms demonstrating industrially relevant advantage over classical computation consistently require at least $10^{12}$ reliable quantum operations~\cite{gidney2021factor,lee2021even,HanerEllipticCurve,Gidney2021faulttolerant,riverlanetera,microteraquop}. Therefore, any error correction scheme needs to reduce the logical error rate to $10^{-12}$ or lower, which will require very large distances. 

The number of physical qubits and the amount of information that needs to be processed by the decoder grow significantly with $d$, whereas the time available to process this information remains constant. As a result, a major challenge for the development of effective decoders is demonstrating a sufficiently fast computation even for large distances, so to avert the backlog problem.

\section{Decoding the surface code}
\label{app:decodesurface}

The core challenge in quantum error correction is to preserve a logical state, known as logical memory~\cite{acharya2023SuppressingNature}. For the surface code, this involves initialising a logical state, performing several rounds of syndrome measurements, and finishing with a logical Pauli measurement. We are concerned with the overall effect of errors on the outcome of the logical measurement. Therefore, the decoding problem is to determine whether the logical measurement has been flipped given the observed syndrome and logical measurements.

The circuit used to measure the parity check operators (\cref{fig:qec-cycle}a) has a syndrome qubit that is reset and measured, single qubit gates, and two-qubit gates that map errors onto the syndrome qubit. In addition to the noise on the data qubits, each of these operations potentially introduce additional noise mechanisms. This is why the parity check operators are measured repeatedly, forming the syndrome. If no errors occur, these measurements produce the same results in consecutive rounds. Therefore, an error is detected when there is a change in the outcome of a measurement from one round to the next. We call these changes in measurement outcomes \emph{defects.}

The syndrome is best represented in a decoding graph (\cref{fig:qec-cycle}b). The vertices of the decoding graph correspond to all possible defects. If an error mechanism triggers two defects, we connect the corresponding vertices by an edge. Some error mechanisms only trigger a single defect e.g.~errors on the data qubits on the boundaries of the lattice. To capture these error mechanisms, we connect the corresponding defects to virtual boundary vertices. By taking the XOR of consecutive rounds of syndrome measurements, we can identify the syndrome with the set of defects that have been triggered. The decoding problem can now be rephrased as determining the most likely logical measurement given the defects in the decoding graph generated by running the syndrome extraction circuit.

The decoding graph is actually two disjoint decoding graphs, one generated by $Z$ checks, to correct $X$ errors, and one generated by $Z$ checks, to correct $X$ errors. We decompose the $Y$ errors included in our noise model into $X$ and $Z$ errors, which are handled in the appropriate decoding graph. More accurate decoding schemes exist that handle $Y$ errors by correlating the two decoding graphs \cite{fowlerCorr2013,acharya2023SuppressingNature}. We save an investigation into hardware implementations of correlated decoders for future work.

\section{Noise model}
\label{app:noise-model}

Noise models vary both in the level of errors they produce as well as the types of error mechanism that can occur, which have a significant impact on the decoder performance. Throughout this work, we sample syndromes using the Clifford circuit simulator Stim~\cite{gidney2021Stim} with several independent noise channels, parameterized by a single probability $p$, that give a rough approximation of noise channels characteristic of a generic superconducting device:
\begin{itemize}
    \item Depolarisation of both qubits after each 2-qubit gate with probability $p$.
    \item Depolarisation of each idle qubit and after each single-qubit gate, including measurement and reset operations, with probability $p/10$.
    \item Randomly change the result of a measurement with probability $p$.
\end{itemize}
We use the parametrisation where depolarising a single qubit means applying a random non-identity Pauli error:
\begin{equation}
    \mathcal E(\rho) = (1-p)\rho + \frac{p}{3}(X\rho X + Y\rho Y + Z\rho Z)
\end{equation}
Depolarising two qubits means applying one of the 15 non-identity two qubit Pauli errors uniformly at random (so that each two qubit error occurs with probability $p/15$).

\begin{figure*}[ht]
    \centering
    \includegraphics[width=0.9\linewidth]{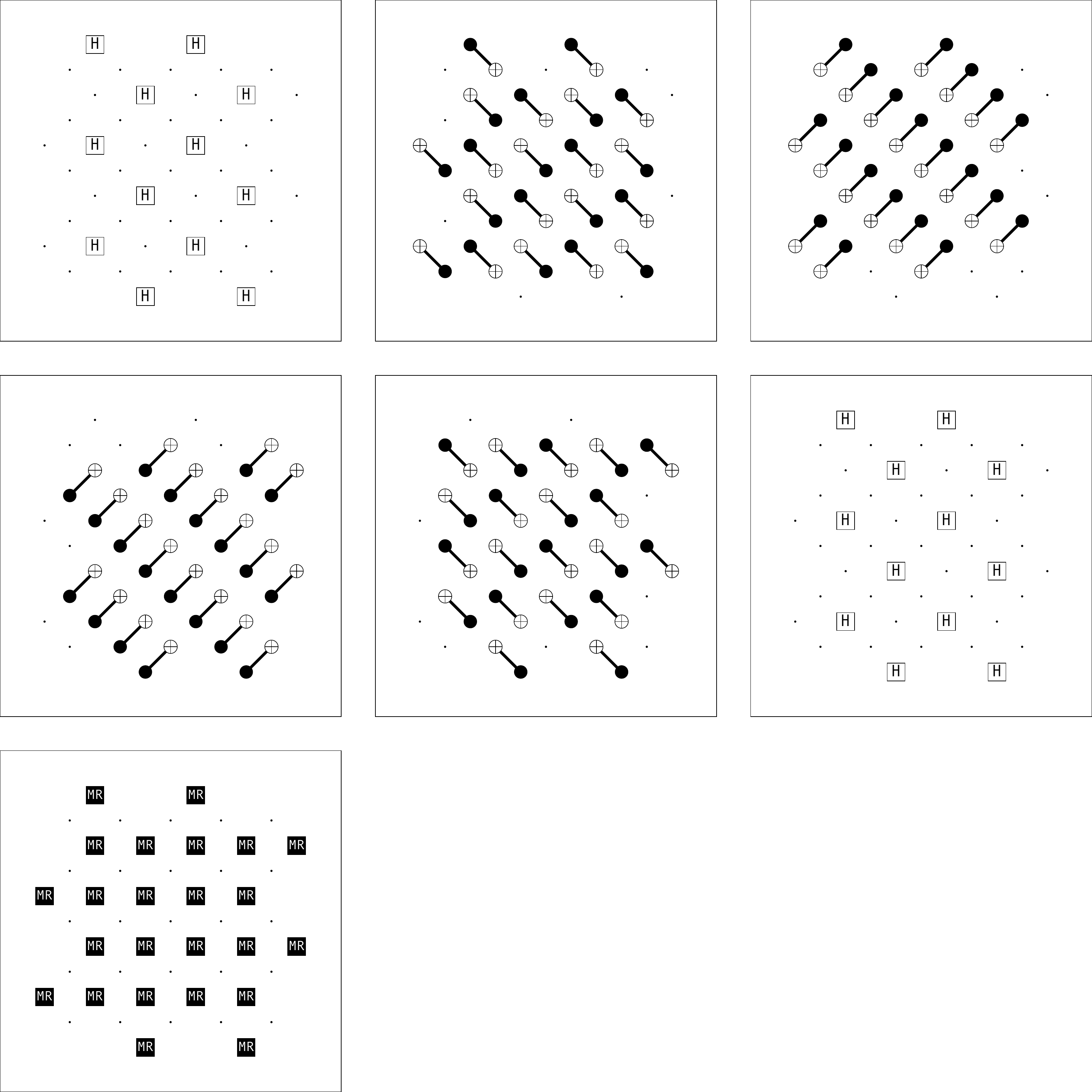}
    \caption{Diagram of the circuit for a round of check measurements for distance 5 rotated planar code. Full circuits together with the noise model are available as stim \cite{gidney2021Stim} files on Zenodo with the DOI identifier \href{https://doi.org/10.5281/zenodo.11621878}{10.5281/zenodo.11621877}.}
    \label{fig:circuit_diagram}
\end{figure*}

Our circuit-level noise model, illustrated in \cref{fig:qec-cycle}c, generates a larger quantity and variety of defects than the phenomenological noise model~\cite{dennisetal08} which abstracts away the details of generating the syndrome, and is a less realistic noise model. On a distance $d=23$ rotated planar surface code with $p=0.1\%$, our noise model produces defects at a rate of $1.35\%$ (or about 3.6 defects per round), while the phenomenological noise model~\cite{liyanage2023scalable} with the same probability $p$ produces defects at a rate of $0.38\%$ (or about 1 defect per round, \cref{fig:qec-cycle}d). The details of the circuit that is used in our experiment are in \cref{fig:circuit_diagram}
 and the stim \cite{gidney2021Stim} circuits used to simulate the noise are available on Zenodo with the DOI identifier \href{https://doi.org/10.5281/zenodo.11621878}{10.5281/zenodo.11621877}.

\section{Application to other codes}
\label{app:other_codes}

In the future, different target applications will require different code distances based on the length of the computation to be performed. For this reason, decoding hardware that can only deal with a very specific code and code distance has limited value in the real world. 

The decoding graph in the CC decoder is encoded by the distance calculation function, and this is the only part of the design that has to change based on the code. This makes CC and its implementations applicable to a large family of codes. The current implementation is optimised for the surface code where efficient closed form distance functions exist. For more general decoding graphs, closed-form distance functions could be developed by using graph embedding techniques, or the design could be adapted to utilise efficient lookup tables of distances.

\section{Coordinate embedding and distance function}
\label{app:coordinate}

\begin{figure*}[ht]
    \centering
    \includegraphics[width=0.9\linewidth]{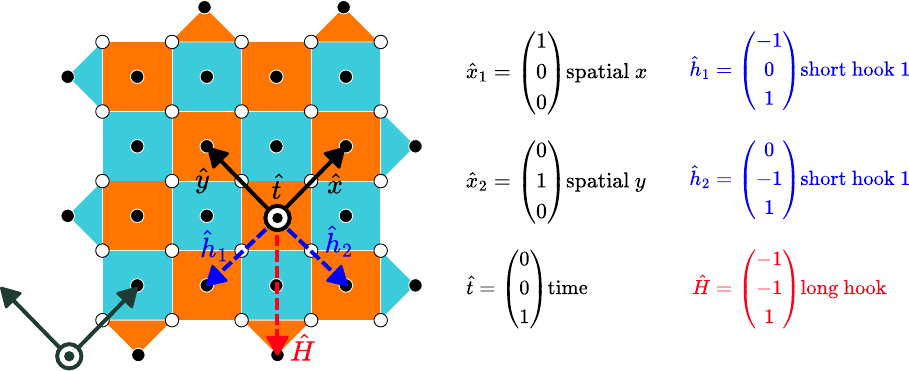}
    \caption{Rotated planar code graph coordinate system embedding. (a) Sketch of the 5x5 rotated planar code and the principal axes. The coordinate system is aligned at 45° to the code. With the phenomenological noise model, graph is described by the nearest-neighbour connections along the $\hat{x}$, $\hat{y}$ and $\hat{t}$ directions. With the circuit-level noise model we get additional diagonal connections that are dependent on the schedule of entangling operations in the circuit and that connect nodes that are both spatially and time-like separated. In the literature, these are often referred to as hooks. Here, they are represented by $\hat{h}_1$, $\hat{h}_2$ and $\hat{H}$. A choice of the origin for the coordinate system as used in CC is shown in the bottom left corner of the figure. }
    \label{fig:coordinate_embedding}
\end{figure*}

The collision clustering algorithm relies on fast and efficient calculation of the shortest path between any two nodes of the decoding graph. This can be done for a large number of cases by isometrically embedding the graph in a normed coordinate space and having a distance function compute the norm between the node coordinates. Here we outline an approach to embedding the rotated planar code graph for both the phenomenological and circuit-level noise model.

Each node can be trivially (not necessarily isometrically) embedded in a 3D space by labelling it with its coordinates:
\begin{equation}
    \label{eqn:embedding}
    X = (x_1, x_2, t)
\end{equation}
First, consider the phenomenological noise model and the unweighted decoding graph. In this case, there are no hook edges and all graph connections are along the principal axes of the coordinate system (see \cref{fig:coordinate_embedding}). The distances between nodes of the graph are then computed by simply taking a Manhattan distance:
\begin{equation}
    D(X_1, X_2) = || X_1 - X_2||_1 = |\Delta x_1| + |\Delta x_2| + |\Delta t|
\end{equation}
We can also easily add weights to such a graph if we assume that the weights obey the translational symmetry (i.e., all edges along a particular direction have the same weight). In this case, we have 3 weights $w_1$, $w_2$, $w_3$ along the $x_1$, $x_2$, $t$ directions respectively. The correct distance can again be calculated with the Manhattan norm, but now the embedded coordinates need to be scaled:
\begin{alignat}{2}
    &X &&= (w_1 x_1, w_2 x_2, w_3 t) \\
    D(X_1, &X_2) &&= || X_1 - X_2||_1 \nonumber\\
    &           &&= w_1|\Delta x_1| + w_2|\Delta x_2| + w_3|\Delta t| \nonumber
\end{alignat}
In the circuit level noise model, hook edges are added which increases the complexity (see \cref{fig:coordinate_embedding}). The diagonal hook edges mean we can not isometrically embed the graph in 3D space. However, the unweighted circuit-level graph can be isometrically embedded into 4D space using the L1 norm:
\begin{alignat}{3}
    \label{eqn:hooked_distance}
    &X &&= &&\left( \frac{x_1}{2}, \frac{x_2}{2}, \frac{x_1 + t}{2}, \frac{x_2 + t}{2} \right) \\
    D(X_1, &X_2) &&= && || X_1 - X_2||_1 \nonumber\\
    & &&= &&\frac{1}{2} \left( |\Delta x_1| + |\Delta x_2|  + \right.\nonumber\\
    & &&  &&\left. +|\Delta x_1 + \Delta t| + |\Delta x_2 + \Delta t| \right)\nonumber
\end{alignat}
In the CC decoder presented in the main paper, the clusters are labelled by their coordinates (\cref{eqn:embedding}) and the distance calculated when needed according to \cref{eqn:hooked_distance}. The boundary is treated as a special node and the distance to the boundary calculated as $\min (x_1, d-x_1)$ where $d$ is code distance and the origin of the coordinate system is as defined in \cref{fig:coordinate_embedding}.
The \cref{eqn:hooked_distance} can be extended to the weighted graph assuming translational symmetry, but requires an embedding in 7D space and is beyond the scope of this paper.

\bibliography{main}

\end{document}